\documentstyle[aps,eqsecnum]{revtex}
\input epsf
\draft
\begin{document}
\twocolumn[\hsize\textwidth\columnwidth\hsize\csname
@twocolumnfalse\endcsname
\title{
Effects of QCD Resummation on $W^+h$ and $t\bar b$ \\
  Production at the Tevatron }
\author{S. Mrenna\cite{mrenna_email}}
\address
{High Energy Physics Division\\
Argonne National Laboratory \\
Argonne, IL  60439}
\author{C.--P. Yuan\cite{yuan_email}}
\address
{Department of Physics and Astronomy\\
Michigan State University \\
East Lansing, MI  48824}
\date{\today}
\maketitle
\begin{abstract}
\noindent
The resummation of multiple soft gluon emission affects the
production rate and kinematic distributions of $W^+h$ 
(where $h$ is a Higgs boson) and $t \bar b$ pairs at
the Tevatron with $\sqrt{s}=2$ TeV.  
Using the Collins--Soper--Sterman resummation
formalism, the production rate is enhanced over the
next--to--leading--order (NLO) prediction by 2--3\% for 
the $W^+h$ process, for Higgs boson masses between
80--120 GeV, and over 3\% for the $t\bar b$ process for
$m_t=175$ GeV.  After resummation, the $t\bar b$ rate
changes by 12--13\% when $m_t$ is varied by $\pm 5$ GeV.
Various kinematic distributions are presented for the
individual final state particles and for the pair.
The explicit radiation of hard gluons in NLO QCD is 
included also for the $t\bar b $ final state.

\end{abstract}
%
%
                                                        \vskip2pc]

\narrowtext
\section{Introduction}
\indent

The Tevatron collider at Fermilab, running at a center--of--mass
energy $\sqrt{s}=2.0$ TeV with a higher luminosity, will produce
a sizeable number of electroweak gauge bosons at large invariant masses,
which can subsequently decay to top quarks $t$ or Higgs bosons $h$.
Studying the properties of $t$ and 
$h$ are necessary to complete our understanding of the Standard Model
(SM).
The process ${\rm p \bar p}\to {W^\pm}^* \to W^\pm h$ is a 
promising avenue for detecting a light Higgs boson, which 
subsequently decays
$h(\to b \bar b)$\cite{higgs}, or constraining the Higgs boson
mass $M_h$.
The process ${\rm p \bar p}\to {W^+}^* \to t \bar b~
({W^-}^* \to \bar t b)$ has been proposed 
to bound the CKM matrix element $V_{tb}$ in a kinematic region where
the parton distribution functions are well known\cite{vtb}.  
After the decay $t\to b W^+ (\bar t \to \bar b W^-)$, 
the second process is also a background to the first process, so it
is interesting to study both in parallel.
To determine the sensitivity of the Tevatron to
$M_h$ and $V_{tb}$ or
to observe any new physics effect in the $W^\pm b\bar b$
system, one has to know first the SM prediction for
the production rate and the kinematic distributions of the
final state system.  The purpose of this study is to quantify the
accuracy of the SM prediction for the next run at the Tevatron.

To test the SM or probe new physics, it is important to study the
kinematics of the final state particles, like $h$ and $t$.
An accurate prediction of the rate and kinematics of particle
production at hadron colliders must include initial and possibly
final state radiation of gluons in QCD.
Electroweak corrections are typically much smaller than higher
order QCD corrections \cite{wnlo}.  
The effects of multiple soft gluon resummation on the kinematics 
of on shell $W^\pm$
bosons and the leptonic decay products have been studied in 
some detail\cite{csaba}.  The situation
is not so complete for off shell $W^\pm$ boson production at large 
invariant mass. 
For the processes of interest,
the initial and final (if any) state QCD corrections at NLO factorize
separately.  The predicted NLO rate for $W^\pm h$, with only
initial state corrections, differs from the leading order (LO) rate
by 36\% for $M_h=$80 GeV.  The rate for $t \bar b$ production, 
with initial and final
state corrections, differs by 60\% for $m_t=$175 GeV.
Initial state soft gluon resummation may affect the size of these corrections
and modify the kinematic distributions of the final state as
expected from a NLO calculation.
Furthermore, to
accurately predict the kinematics of the $t\bar b$ final state at NLO,
explicit hard gluon radiation must be included.

Sec.~\ref{sec2} contains a review of the 
Collins--Soper--Sterman (CSS) resummation formalism used in this study. 
We closely follow the notation used 
in Ref.~\cite{sterman}.
In Sec.~\ref{sec3}, we
present our numerical results for 
$q \bar q' \rightarrow {W^+}^*\to W^+h$ 
production using this formalism.  
We show how to incorporate the effects of final state
radiation in Sec.~\ref{sec4}, and present numerical results
for $q \bar q' \rightarrow {W^+}^*\to t\bar b$.
Finally, Sec.~\ref{sec5} contains our conclusions.


\section{The CSS Resummation Formalism}
\label{sec2}
\indent

Soft gluon resummation has been applied successfully to predict
the rate and kinematics of electroweak gauge boson production at hadron
colliders \cite{sterman,altarelli,collins,arnold}.  For $W^\pm$
bosons, the CSS formalism has been applied only to leptonic final states. 
The same formalism can be directly applied to 
${W^+}^*\to W^+h$ production.  However, for ${W^+}^*\to t\bar b$
production, there are also singularities associated with the final state
radiation which need to be handled separately.

The resummed differential cross section 
in the CSS formalism for the production of gauge boson $V$ at a hadron
collider with a center--of--mass energy $\sqrt{s}$ is
  \begin{eqnarray} 
  \left( { d \sigma(h_1 h_2 \rightarrow V(\to d_1d_2)X)
  \over dQ^2 \, dy \, dQ^2_T \, d\phi_V \, d\cos{\theta} \,
  d\phi}
  \right)_{\rm resum} = 
  \nonumber \\ 
  { \beta \over 96 \pi^2 s }{Q^2 \over (Q^2-M^2)^2+M^2\Gamma^2} \, 
  \bigg\{ 
  \int_{}^{} {d^2 b \over (2 \pi)^2}  \, e^{i {\vec Q_T} \cdot {\vec b}} \, 
  \nonumber \\
  \times \sum_{j,k}{\widetilde{W}_{jk} 
   (b_*,Q,x_1,x_2, \theta, \phi,C_1,C_2,C_3)} \, 
  \nonumber \\
  \times F^{\rm NP}_{jk} (b,Q,x_1,x_2) 
  + ~  Y(Q_T,Q,x_1,x_2, \theta, \phi,C_4) \bigg\},
  \label{ResFor}
  \end{eqnarray}
where $Q, M, \Gamma, y, Q_T$ and $\phi_V$ are the mass, on shell
mass, width, rapidity, transverse
momentum, and azimuthal angle of $V$ in the laboratory frame,
$\theta$ and $\phi$ are the polar and azimuthal angle
of the final state particle $d_1$ [see Eq.~(\ref{bigeqn})]
in the Collins--Soper frame\cite{CSFrame}, 
$x_1 = ({Q/\sqrt{s}})\exp(y), x_2 = ({Q/\sqrt{s}})\exp({-y})$,
and $\beta=2p^*/Q$, 
where $p^*$ is the momentum of $d_1$ and $d_2$ in the $V$ rest frame.
The renormalization group invariant $\widetilde{W}_{jk}$
is given by
  \begin{eqnarray} 
&  \widetilde{W}_{jk} (b,Q,x_1,x_2, \theta, \phi,C_1,C_2,C_3) =  
&  \nonumber \\ 
&  \exp \left\{ -S(b,Q,C_1,C_2) \right\} |V_{jk}|^2  
&  \nonumber \\ 
& \times \bigg\{
  \left[  
  \left( C_{jl} \otimes f_{l/h_1} \right) (x_1) ~
  \left( C_{km} \otimes f_{m/h_2} \right) (x_2) \right. 
&  \nonumber \\ 
&  \left.
  +\left( C_{kl} \otimes f_{l/h_1} \right) (x_1) ~
  \left( C_{jm} \otimes f_{m/h_2} \right) (x_2) \right]  
&  \nonumber \\ 
&  \times g_L^2 f_S^2{\cal H}^S_{jk}(Q,\cos\theta,m_{d_1},m_{d_2}) 
&  \nonumber \\ 
&  + \left[  
  \left( C_{jl} \otimes f_{l/h_1} \right) (x_1) ~
  \left( C_{km} \otimes f_{m/h_2} \right) (x_2) \right. 
&  \nonumber \\ 
&  \left.
 - \left( C_{kl} \otimes f_{l/h_1} \right) (x_1) ~
  \left( C_{jm} \otimes f_{m/h_2} \right) (x_2) \right]  
&  \nonumber \\ 
&  \times g_L^2 f_A^2{\cal H}^A_{jk}(Q,\cos\theta,m_{d_1},m_{d_2}) \bigg\},
  \label{WTwi}
&  \end{eqnarray}
where $V_{jk}$ is a CKM matrix element for the initial state partons
$j$ and $k$, $S$ is a Sudakov form
factor, 
$m_{d_1}, m_{d_2}$ are the final state masses,
${\cal H}^{S(A)}$ is a symmetric (antisymmetric) function
in $\cos\theta$,
$g_L^2 = G_F
M_W^2/\sqrt{2}$, $f_{S(A)}$ is a coupling for the symmetric
(antisymmetric) function, 
and $\otimes$ denotes the convolution integral of the Wilson
coefficients $C$
and the parton distribution functions (PDF's)
  \begin{eqnarray} 
  \left( C_{jl} \otimes f_{l/h_1} \right) (x_1) =  
  \nonumber \\
  \int_{x_1}^{1} {d \xi_1 \over \xi_1} \, 
  C_{jl}( {x_1 \over \xi_1}, b, \mu={C_3 \over b}, C_1,C_2)
  \nonumber \\
  \times f_{l/h_1}(\xi_1, \mu={C_3 \over b}).
  \label{Convol}
  \end{eqnarray}
For the processes under consideration, $j=q, k=\bar q^{'}$,
excluding $t$ quarks.
The dummy indices $l$ and $m$ 
sum over quarks and anti-quarks or gluons, and
summation on double indices is implied.
The angular function ${\cal H}^{S(A)}$ and
coupling constants $f_{S(A)}$ in 
Eq.~(\ref{WTwi}) for the 
$d_1=W^+,~d_2=h$ final state are
  \begin{eqnarray} 
&    
   {(M_W^2/Q^2)} \left [ 2 + {(p^*/M_W)}^2 (1 - \cos^2\theta) \right], & (S)
  \nonumber \\
&       0, & (A)
  \nonumber \\
&      f_S^2 = {\sqrt{2} G_F M_W^2}, f_A^2=0, &
  \nonumber
  \end{eqnarray}
and for the 
$d_1=t,~d_2=\bar b$ final state are
  \begin{eqnarray} 
&       3\left [ (1-(m_t^2-m_b^2)^2/Q^4+{(2p^*/Q)}^2\cos^2\theta) \right],& (S)
       \nonumber \\
&       3\left [ {(4p^*/Q)}\cos\theta \right], & (A)
       \nonumber \\
&       f_S^2=f_A^2=g_L^2|V_{tb}|^2. &
   \nonumber 
  \end{eqnarray}
The Sudakov form factor $S(b,Q,C_1,C_2)$ in Eq.~({\ref{ResFor}) is defined as
  \begin{eqnarray} 
  S(b,Q,C_1,C_2) =  
  \nonumber \\
  \int_{C_1^2/b^2}^{C_2^2Q^2}
  {d {\bar \mu}^2\over {\bar \mu}^2}
       \left[ \ln\left({C_2^2Q^2\over {\bar \mu}^2}\right)
        A\big(\alpha_s({\bar \mu})\big) +
        B\big(\alpha_s({\bar \mu})\big)
       \right],
  \label{SudExp}
  \end{eqnarray}
where $\alpha_s$ is the strong coupling constant,
and the functions $A$, $B$ and the  Wilson coefficients $C_{jl}$  
were given in Ref.~\cite{sterman}.
The constants 
$C_1$, $C_2$ and $C_3 \equiv \mu b$ arise
when solving the renormalization group equation for the 
$\widetilde{W}_{jk}$.
The canonical choice of these renormalization constants is
$C_1 = C_3 = 2 \exp({-\gamma_E}) \equiv b_0 $ and $C_2 = 1$, where
$\gamma_E$ is the Euler constant.  In general,
the choice
$C_1=C_2 b_0$ and $C_3=b_0$
eliminates large constant factors in the
expressions for the $A, B$ and $C_{jk}$ functions.
As shown in Eq.~(\ref{SudExp}), the upper limit of the integral
for calculating the Sudakov factor is ${\bar \mu}=C_2 Q$, which sets
the scale of the hard scattering process when evaluating
the renormalization group invariant
quantity $\widetilde{W}_{jk}$, as defined in Eq.~(\ref{WTwi}).
The lower limit ${\bar \mu}\equiv C_1/b = b_0/b$ determines the onset of
nonperturbative physics.  When integrated over the impact parameter
$b$, the $\widetilde W$ term is referred to as the CSS piece.

The $Y$ piece in Eq.~(\ref{ResFor}) is defined as
  \begin{eqnarray}
  Y(Q_T,Q,x_1,x_2,\theta,\phi,C_4) =  
  \nonumber \\
  \int_{x_1}^{1} {d \xi_1 \over \xi_1}
  \int_{x_2}^{1} {d \xi_2 \over \xi_2}
  \sum_{N=1}^{\infty} \left[{\alpha_s(C_4 Q) \over \pi} \right]^{N} 
  \nonumber \\
  \times f_{l/h_1}(\xi_1;C_4 Q) \, R_{lm}^{(N)} (Q_T,Q,{ x_1 \over \xi_1},
  { x_2 \over \xi_2},\theta,\phi)   
  \nonumber \\
  \, \times f_{m/h_2}(\xi_2;C_4 Q), 
  \label{RegPc}
  \end{eqnarray}
where the functions $R_{lm}^{(N)}$ 
only contain contributions less singular than 
${Q_T^{-2}} \times(1 \, {\rm or} \, \ln({Q^2 /Q_T^2}))$ 
as $Q_T \rightarrow 0$ in the fixed order perturbative calculation.
The perturbative expansion is optimized by choosing $C_4=1$, so that
the scale of the $Y$ piece is $Q$.

In Eq.~(\ref{ResFor}), the impact parameter $b$ is to be integrated 
from 0 to $\infty$. 
However, for $b \ge b_{\rm max}$, which corresponds to an energy scale 
much less than $1/b_{\rm max}$, the 
QCD coupling $\alpha_s$ becomes so large that a perturbative 
calculation is no longer reliable.
The nonperturbative physics in this region is 
described by an empirically fit function 
$F^{\rm NP}$ with the general
structure
  \begin{eqnarray}
  F^{\rm NP}_{jk} (b,Q,Q_0,x_1,x_2) = 
  \nonumber \\
  \exp \bigg\{-\ln \left( Q^2\over Q^2_0 \right) h_1(b)
    -h_{j/h_1}(x_1,b) 
  \nonumber \\ 
 -h_{k/h_2}(x_2,b)\bigg\}. 
  \label{FNPh}
  \end{eqnarray}
The functions $h_1$, $h_{j/h_1}$ and $h_{k/h_2}$ cannot be calculated using 
perturbation theory and must be measured experimentally.
Furthermore, $\widetilde{W}$ is evaluated at $b_*$,
instead of $b$, with
  \begin{eqnarray}
  b_* = {b \over \sqrt{1+(b/b_{\rm max})^2} }
  \label{bStar}
  \end{eqnarray}
such that $b_*$ never exceeds $b_{\rm max}$.

To obtain the kinematics of the final state products $d_1$ and $d_2$ from
$V^*\to d_1 d_2$, we transform the four momentum
of $d_1$ ($\equiv p^\mu$) and $d_2$ 
($\equiv {\bar p}^\mu$) from the Collins--Soper frame to the laboratory frame.
The resulting expressions are 
  \begin{eqnarray} & &
  p^\mu = {Q \over 2}( {q^\mu \over Q} + \sin\theta\cos\phi
X^\mu + \sin\theta\sin\phi Y^\mu + \cos\theta Z^\mu), \nonumber \\
 & & {\bar p}^\mu = q^\mu - p^\mu, \nonumber \\
 & &  q^\mu = (M_T\cosh y,Q_T \cos\phi_V, Q_T \sin\phi_V, M_T\sinh y),
\nonumber \\
 & &  X^\mu = -{Q \over Q_T M_T}(Q_{+}n^\mu + Q_{-}{\bar n}^\mu - 
{M_T^2 \over Q^2}q^\mu), \nonumber \\
 & & Y^\mu = \epsilon^{\mu\nu\alpha\beta}{q_\nu \over Q}Z_\alpha X_\beta,
\nonumber \\
 & & Z^\mu = {1 \over M_T}(Q_{+}n^\mu - Q_{-}{\bar n}^\mu),
\label{bigeqn}
  \end{eqnarray}
with $q^\mu = (q^0,q^1,q^2,q^3)$, 
$Q_\pm = (q^0\pm q^3)/\sqrt{2}$, $Q=\sqrt{q^2}$, 
$M_T = \sqrt{Q^2+Q_T^2}$, $y=\ln({Q_{+}/Q_{-}})/2$,
$n^\nu = (1,0,0,1)/\sqrt{2},$ ${\bar
n}^\nu= (1,0,0,-1)/\sqrt{2}$, and
$\epsilon^{0123}=-1$.
In the Collins--Soper frame of reference, the $W^*$ is at rest and the
z axis is defined as the bisector of the angle formed by $\vec{p}_{h_1}$
and $-\vec{p}_{h_2}$, where $\vec{p}_{h_1}$ and $\vec{p}_{h_2}$ 
are the momentum of 
the initial state
hadrons.  As suggested by the notation, $X^\mu,Y^\mu,Z^\mu$ are the 
transformations of the $x-,y-,z-$ components of $d_1$, respectively, from
the CSS to the laboratory frame.  Similar expressions can be derived
when there are more particles in the final state as discussed in
Sec.~\ref{sec4}.

\section{The Numerical Results of Initial State Resummation}
\label{sec3}
\indent

In this section, we present numerical results for the 
$q \bar q' \rightarrow {W^+}^*\to W^+ h$ process
after applying the resummation formalism outlined in the previous
section at  
the upgraded Tevatron
with $\sqrt{S}=2.0$\,TeV.  
For these results, we have assumed $m_t$ = 175\,GeV and
$m_b$ = 5\,GeV 
and several values of
the Higgs boson mass $M_h$.  In the 
following, distributions will be shown for ${W^+}h$
production only.
This allows us to exhibit asymmetries along the beam axis.  The
distributions for ${W^-}h$ production can be obtained by reflection
about rapidity $y=0$. 

As explained in the previous section, 
the CSS piece depends on the renormalization 
constants $C_1, C_2=C_1/b_0$ and $C_3=b_0$, as well as a few
other implicit parameters.
The choice of $C_2$ indicates that the hard scale of the
process is $C_2 Q$.
For this study, we use $C_2=1$, which fixes the scale at the off shell
gauge boson mass.
We also use 
the CTEQ3M NLO PDF's \cite{cteq3m}, 
the
NLO expression for $\alpha_s$, and
the non-perturbative function~\cite{glenn}
  \begin{eqnarray}
  F^{\rm NP} (b,Q,Q_0,x_1,x_2)  = 
  \nonumber \\  
 \exp \bigg\{- g_1 b^2 - g_2 b^2 \ln\left( {Q \over 2 Q_0} \right) 
  \nonumber \\
  -g_1 g_3 b \ln{(100 x_1 x_2)} \bigg\}, 
  \label{FNPg}
  \end{eqnarray}
where $g_1 = 0.11\,{\rm GeV}^2$, 
$g_2 = 0.58\,{\rm GeV}^2$, $g_3 = -1.5\,{\rm GeV}^{-1}$ 
and $Q_0 = 1.6\,{\rm GeV}$.
The choice $b_{\rm max} = 0.5~{\rm GeV}^{-1}$ 
was used in this fit.
\footnote{
These values were fit for CTEQ2M PDF and $C_2$=1, 
and in principle should be refit for CTEQ3M PDF.}
Finally, the
CSS piece is fixed by specifying the order in $\alpha_s$ of the
$A,B$ and $C_{jk}$ functions.  We adopt the notation $(M,N)$ to represent
the order in $\alpha_s$ of $A^{(M)}, B^{(M)}$ and $C^{(N)}_{jk}$.
The choice $(1,0)$, for
example, means that $A$ and $B$ are calculated to order $\alpha_s$,
while $C_{jk}(z)$ is either 0 or $\delta(1-z)$ depending on $j$ and $k$.  

The CSS piece alone gives an accurate description of the kinematics
of the off shell $W^+$ boson for $Q_T \ll Q$.  However, because of the
soft gluon approximations made in the resummation, the CSS piece becomes
negative when $Q_T \sim Q$.  The regular $Y$ piece accounts for
terms neglected in $\widetilde W$ at any fixed order in perturbation
theory.  The sum CSS + $Y$ can become
negative for $Q_T\sim Q$, but, when integrated over $Q_T$ from 0 to $Q$,
returns the NLO result for the total cross section and $Q$ distribution.
We have checked this agreement numerically.  Since the $Q_T$ distribution
is a physical observable, negative differential cross sections are not
allowed.  Therefore, when CSS + $Y$ no longer gives an accurate description
of the physics, one should use a prediction at a fixed order in 
$\alpha_s$, since the perturbation series in $\alpha_s$ gives a better
convergence for $Q_T \sim Q$.  Switching between
the two formalisms requires a matching prescription.  Our prescription is
to switch smoothly to the fixed order differential cross section once
CSS+$Y$ becomes smaller than the fixed order prediction as 
$Q_T\simeq Q$.
The cross section after matching, then, gives a different prediction than
the NLO result.  We refer to a full resummed calculation, including the
$Y$ piece and matching, as CSS(M,N), where (M,N) specify the order of
the $A,B,$ and $C$ functions.

Part of the next--to--next--to--leader--order (NNLO) 
prediction for $W^*$ production can be included through
$A^{(2)}$ and $B^{(2)}$
in Eq.~(\ref{SudExp})\cite{sterman}.
When matching to the CSS(1,1) (CSS(2,1)) calculation, 
we use the NLO (NNLO) perturbative prediction for high
transverse momentum $W^\pm$ boson production\cite{kauffman}.
We apply the results of our calculations to study
the total rate and
the kinematic distributions of the $W^+h$ pair and the
individual $W^+$ or $h$ produced in hadron collisions.
These do not include any final state effects.

The effects of resummation on the total production rate are illustrated
in Table~\ref{wph_rate}, which compares LO, NLO, the CSS formalism
with $A,B,$ and $C$ functions of order (1,1) and matching, and the CSS
formalism of order (2,1) and matching.  The LO result uses the
CTEQ3L PDF's.  The CSS(1,1) and CSS(2,1)  
predictions are indistinguishable once
matching is used, and reveal a slight enhancement over the NLO
prediction.  We interpret the difference between the NLO and CSS
predictions as a gauge of the intrinsic theoretically uncertainty
in the total rate up to order $\alpha_s^2$.  Referring to
Table~\ref{wph_rate}, higher order corrections to the NLO 
prediction could be as large as 2--3\%.
There will also be some uncertainty from varying the PDF, but we
have not investigated this.

The effect of resummation on various kinematic distributions are
displayed in Figs.~\ref{fig_one}--\ref{fig_five}.  Distributions
for the individual $W^+$ and $h$ are basically changed in rate, 
but not in shape, compared to the LO and NLO predictions 
(see Figs.~\ref{fig_two}--\ref{fig_four}).  On the scale of
the figures, there is not a significant difference between
CSS(2,1) and CSS(1,1), so we do not include CSS(1,1) in these
figures.  However,
those observables which depend on the kinematics of the pair are
sensitive to the order of resummation.  For the purposes of generating
NLO distributions, a $Q_T$ cut of .8 GeV is imposed to separate the
real emission from the virtual corrections.  
$Q_T$ is the transverse momentum of the virtual ${W^+}^*$ boson, or,
equivalently, the $W^+ h$ pair, since we do not include any
final state corrections.
Near this cut, we have little confidence in the ability of the NLO
calculation to accurately describe the $Q_T$ distribution.
Even far from this cut, however,
the NLO $Q_T$ distribution grossly underestimates the $Q_T$ distribution
as expected from resummation (see Fig.~\ref{fig_one}).  
At high $Q_T$,
CSS(2,1) predicts an enhancement over CSS(1,1), which means more hard
gluon radiation.  
Also, CSS(2,1) predicts a large azimuthal angle separation
($\Delta\phi^{W^+h}$) between $W^+$ and $h$ (see Fig.~\ref{fig_five}).

\section{Initial State Resummation With Explicit Final State Radiation}
\label{sec4}
\indent
Using the same approach discussed in Section \ref{sec3}, we have
also studied the effects of soft gluon resummation 
on the process
$q \bar q' \rightarrow {W^+}^*\to t \bar b$
for several values
of $m_t$ = 170, 175, 180 \,GeV.  
The presence of hadronic final states, however, adds a complication.
In Ref.\cite{ttresum}, the CSS formalism was applied to the production of
$t \bar t$ pairs at the Tevatron.  The effects of final state radiation
were incorporated in this result by resumming all of the final
state logarithms that behave like $Q_T^{-2}\ln (Q_T/Q)^2$ and adding
these coherently to those similar terms from the initial state.
This was justified on the grounds that color links the initial state
to the final state at NLO. Also,
previous studies of top quark decay at NLO show that most gluon
radiation occurs off the $b$ quark line, so that final state radiation
off the $t$ quark line is a smaller effect\cite{qcdrad}.
For the production of a hadronic final state through an s channel
$W^*$ boson, there is no interference at NLO between Feynman diagrams
with initial and final state QCD corrections, so that the effect
of initial state corrections can be calculated separately from final
state corrections.
To calculate the total cross section for $W^*\to t\bar b$, we resum
the initial state multiple soft gluon radiation and include
the effect of QCD radiation at NLO for hadronic final
states by multiplying the production rate for
${\rm p \bar p}\to {W^+}^* \to t \bar b~$ by a factor of 
$K_{\rm fsr}(Q^2)$, a function
of only $Q^2$ once the masses of $t$ and $\bar b$ are fixed~\cite{wkfactor}.  

The effects of initial state resummation with the final state
QCD correction $K_{\rm fsr}$ on the total production rate
are summarized in
Table~\ref{tbb_rate}, which compares the LO, NLO, CSS(1,1), and CSS(2,1) 
predictions for various values of $m_t$.  
For $m_t=175$ GeV, the enhancement from the CSS
formalism over NLO is about 3.5\%, which is an estimate of 
the effects of higher order corrections.  The enhancement over
the LO result is 66\%.
As a comparison, 
the variation in rate from
a change in $m_t$ by $\pm$ 5 GeV around a nominal value of 175 GeV 
is 12--13\%.  Our NLO correction from initial state radiation, which is
37\% for $m_t = 175$ GeV, agrees
with Ref.~\cite{wnlo}.  However, our NLO correction from 
final state radiation, which is an
additional 17\% enhancement, is about 3.4\% higher than
Ref.~\cite{wnlo}.  The enhancement of the
resummed result over the NLO result is based on {\it our} calculation
of the NLO rate.
A study of the kinematic distributions of the final state particles
$t$ and $\bar b$ at NLO in QCD
requires a separation of the states with and
without explicit hard gluon radiation.
While $K_{\rm fsr}$ can 
predict the change in rate from final state radiation, it
has integrated out the kinematics of the final state particles.  
To remedy this situation, so that we can study the explicit final
states $t\bar b$ and $t\bar b g$,
we calculate the correction function 
$K_{\rm fsr}^{(3)}(Q^2,E_g^{\rm min})$ for
the exclusive process ${W^+}^*\to t\bar b g$ using
a minimum gluon energy $E_g^{\rm min}$ as a regulator.  
The effect of
the regulator on the exclusive process ${W^+}^*\to t\bar b$ can
be calculated using $K_{\rm fsr}^{(2)}(Q^2,E_g^{\rm min})=K_{\rm fsr}(Q^2)-
K_{\rm fsr}^{(3)}(Q^2,E_g^{\rm min})$.  
This is a practical method as long as
the value of $E_g^{\rm min}$ transformed to the laboratory frame
does not exceed the lowest jet energy used by an experiment
in an analysis.  In our numerical results, $E_g^{\rm min} = 1$ GeV.
\footnote{
As discussed in Ref.~\cite{csaba}, if the kinematic cuts imposed to 
select the signal events are correlated with the transverse momentum $Q_T$ of the 
virtual $W$ boson, then the distributions predicted by the NLO and the resummed 
calculations will be different. The difference is the largest near the 
boundary of the phase space, such as in the large rapidity region. }

Figs.~\ref{fig_six}--\ref{fig_eleven} show the effects of
initial state resummation and final state NLO QCD radiation
on the kinematic distributions of the $t$, $\bar b$,
and the pair $t\bar b$.  
The $Q_T$
distribution in Fig.~\ref{fig_six} is still the transverse momentum
of the virtual ${W^+}^*$ boson, so it does not depend on
the separation of final states with and without hard gluon radiation.
For this study, the $t$ and $\bar b$
are produced on shell and treated at the parton level.
No jet definition is used to combine the hard gluon with
$t$ or $\bar b$ partons.
We present results for the resummation calculations 
CSS(2,1).  On the scale of these figures, there is no significant
difference between CSS(2,1) and CSS(1,1).
For comparison, we show the distribution
expected using two body kinematics for $t\bar b$ but still using
the full final state correction $K_{\rm fsr}$.  
The difference between this distribution, denoted as 2--body in the figures, 
and the resummed curve indicates the size of the ``hard'' gluon effects,
that is, the contribution from $t\bar b g$ for a fixed $E_g^{\rm min}$.
When we feel a comparison is useful, we also
include the LO prediction.
For the
kinematics of the individual $t$ or $\bar b$, the effect on the
shape of distributions from final state radiation is important, more
so than from the resummation (see
Figs.~\ref{fig_seven}--\ref{fig_eleven}), and we do not see a 
noticeable difference between the NLO and resummed curves with
our choice of $E_g^{\rm min}$.
The kinematics of the 
$t \bar b$ pair, however, are sensitive to the initial state gluon radiation
(see Figs.~\ref{fig_six} and \ref{fig_twelve}).  
This can be understood as follows.
Similar to the $W^+h$ final state,
intial state gluon radiation only affected distributions of the 
$t \bar b$ pair.
Why is this so?  This is mainly determined by the kinematics 
and the nature of the radiation divergence for the initial and final
states.
In the initial state, the partons are collinear with the beam axis,
and the leading divergence is both soft and collinear.
As a result, the $Q_T$ distribution (see Fig.~\ref{fig_six}) peaks around 
a few GeV, while the rapidity of the pair is very central.
The LO distribution for $P_T^t$, in contrast, peaks near 30 GeV.  Therefore,
there is little effect of the $Q_T$ smearing on $P_T^t$.  The same
argument can be applied to the $t$ rapidity distribution $y^t$.
The final state radiation, on the other hand,
only produces soft divergences, and the final state particles tend to
be central and not collinear with the beam.  Therefore, distributions
with respect to the beam axis, such as $P_T^t$ and
$P_T^b$, are more sensitive to the final state
effects.

\section{Conclusions}
\label{sec5}
\indent

We have studied the effects of QCD resummation in the CSS formalism
on the production of ${W^+}^*$ bosons at the
Tevatron and the kinematics of the final state products $W^+h$ and
$t\bar b$.
Understanding the $W^+h$ production rate is important for quantifying
the Higgs boson discovery potential of the Tevatron.  We find an increase in
rate over the LO result of 40\% for $M_h=80$ GeV, and 29\% for $M_h=120$ GeV.  
This is
only slightly higher than the NLO prediction, even though we have 
included part of the NNLO correction.  Therefore, we estimate the
effect of higher order QCD corrections to be 2--3\% over the NLO
prediction.
The shape of kinematic distributions of the 
$W^+$ and $h$ are not significantly altered from the LO or NLO
predictions.  The kinematics of the $W^+h$ pair, however, can
only be accurately predicted using a resummed calculation, 
and might be 
useful to distinguish signal from background.

Understanding the $t\bar b$ production rate is important for the
determination of the CKM matrix element $V_{tb}$.
We have included the explicit radiation
of final state gluons at NLO, so that we predict the kinematics of
$t \bar b$ and $t \bar b g$ events.    
We find an increase in
rate over the LO result of 66\% for $m_t=175$ GeV, and we
estimate the effect of higher order corrections to be more 
than 3\% over the NLO result.
The total production rate
changes 12--13\% when $m_t$ is varied by $\pm 5$ GeV.  The effects of
final state radiation are more important than 
initial state resummation for
the kinematics of the individual $t$ or $\bar b$.  With explicit gluon
radiation, the $P_T$ spectrum of both $t$ and $\bar b$ are harder, so
that the $b$ tagging efficiency should be higher.
The kinematics of the $t\bar b$ pair are also interesting
observables.

\acknowledgements

We thank C.~Balazs for
discussions.
This work was supported in part by NSF grant PHY-9507683
and by DOE grant W--31--109--ENG--38.

%
\begin{table}
\begin{tabular}[bht]{|c||c|c|c|}\hline
           & \multicolumn{3}{c}{Higgs Mass (GeV)}  \\ \cline{2-4}
Order      & 80       & 100      & 120     \\  \hline
CSS (2,1)  & .305     & .154     &  .083   \\ 
CSS (1,1)  & .304     & .154     &  .083   \\ 
NLO        & .296     & .149     &  .081   \\
LO         & .218     & .109     &  .059   \\
\end{tabular}
\caption{Total cross section in pb for $W^+h$ 
production at the Tevatron in various approximations for
$M_h$=80,100, and 120 GeV.}
\label{wph_rate}
\end{table}
\begin{table}
\begin{tabular}[bht]{|c||c|c|c|}\hline
           & \multicolumn{3}{c}{Top Mass (GeV)}  \\ \cline{2-4}
Order      & 170      & 175      & 180     \\  \hline
CSS (2,1)  & .541     & .478     &  .423   \\ 
CSS (1,1)  & .541     & .477     &  .423   \\ 
NLO        & .523     & .462     &  .409   \\
LO         & .327     & .288     &  .255   \\
\end{tabular}
\caption{Total cross section in pb for $t\bar b$ 
production at the Tevatron in various approximations for
$m_t$=170,175, and 180 GeV.}
\label{tbb_rate}
\end{table}
%
%
\begin{figure}
\centering
\hspace*{0in}
\epsfxsize=3.0in
\epsffile{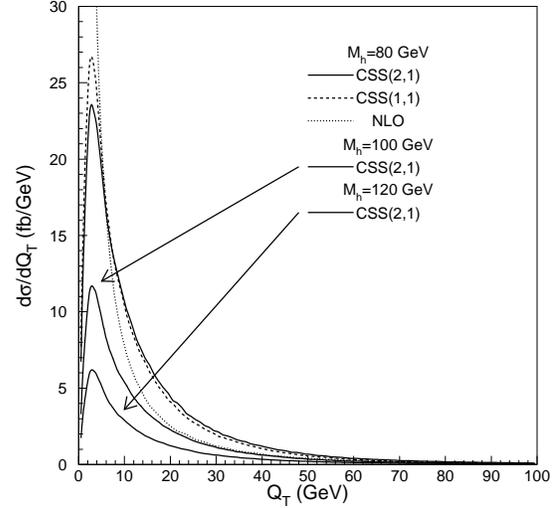}
\vspace*{0in}
\caption{The $Q_T$ spectrum for $W^+h$ Events}
\label{fig_one}
\end{figure}
\begin{figure}
\centering
\hspace*{0in}
\epsfxsize=3.0in
\epsffile{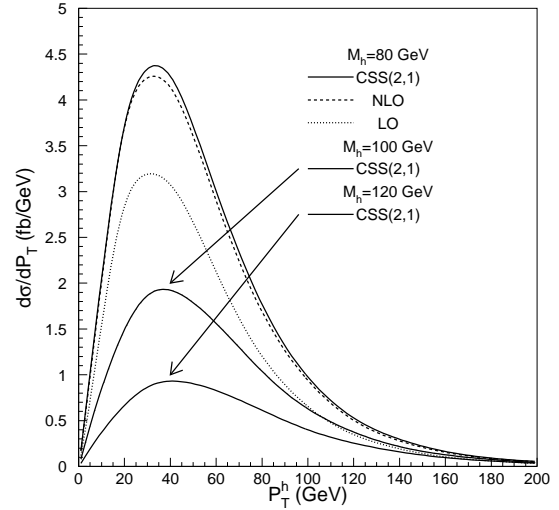}
\vspace*{0in}
\caption{The $P_T^h$ spectrum for $W^+h$ Events}
\label{fig_two}
\end{figure}
\begin{figure}
\centering
\hspace*{0in}
\epsfxsize=3.0in
\epsffile{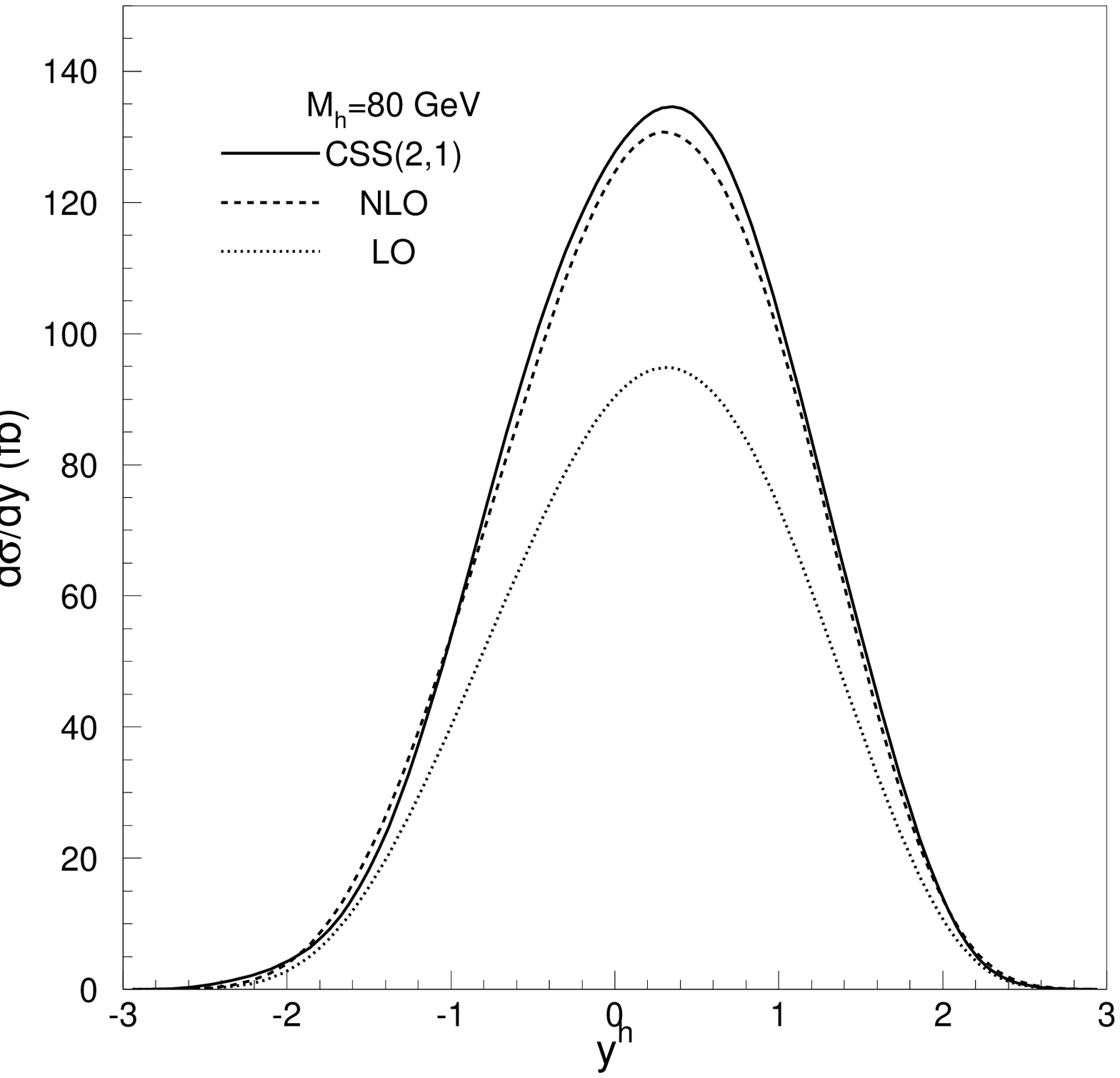}
\vspace*{0in}
\caption{The $y^h$ spectrum for $W^+h$ Events}
\label{fig_three}
\end{figure}
\begin{figure}
\centering
\hspace*{0in}
\epsfxsize=3.0in
\epsffile{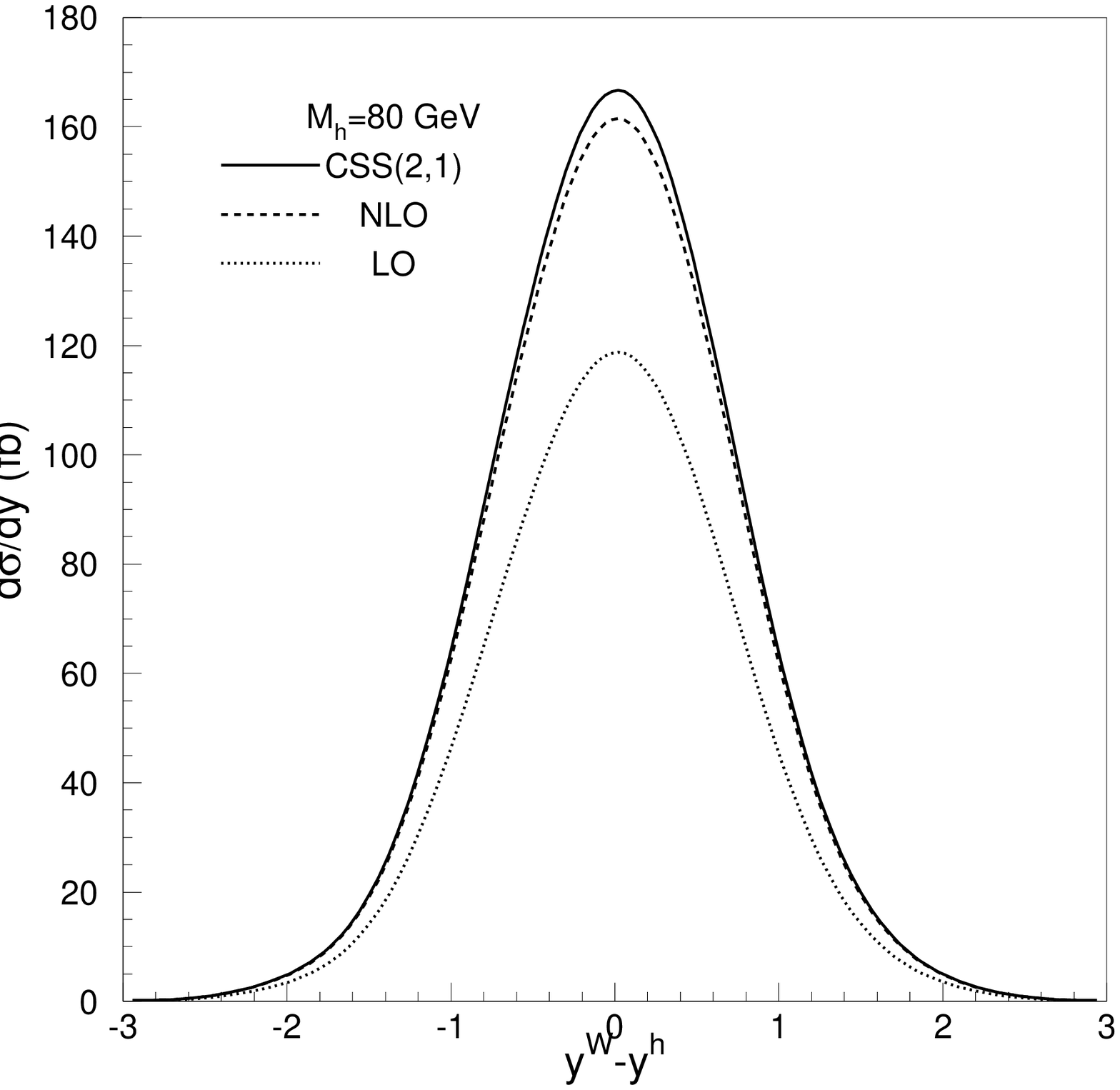}
\vspace*{0in}
\caption{The $\Delta y^{W^+h}$ spectrum for $W^+h$ Events}
\label{fig_four}
\end{figure}
\begin{figure}
\centering
\hspace*{0in}
\epsfxsize=3.0in
\epsffile{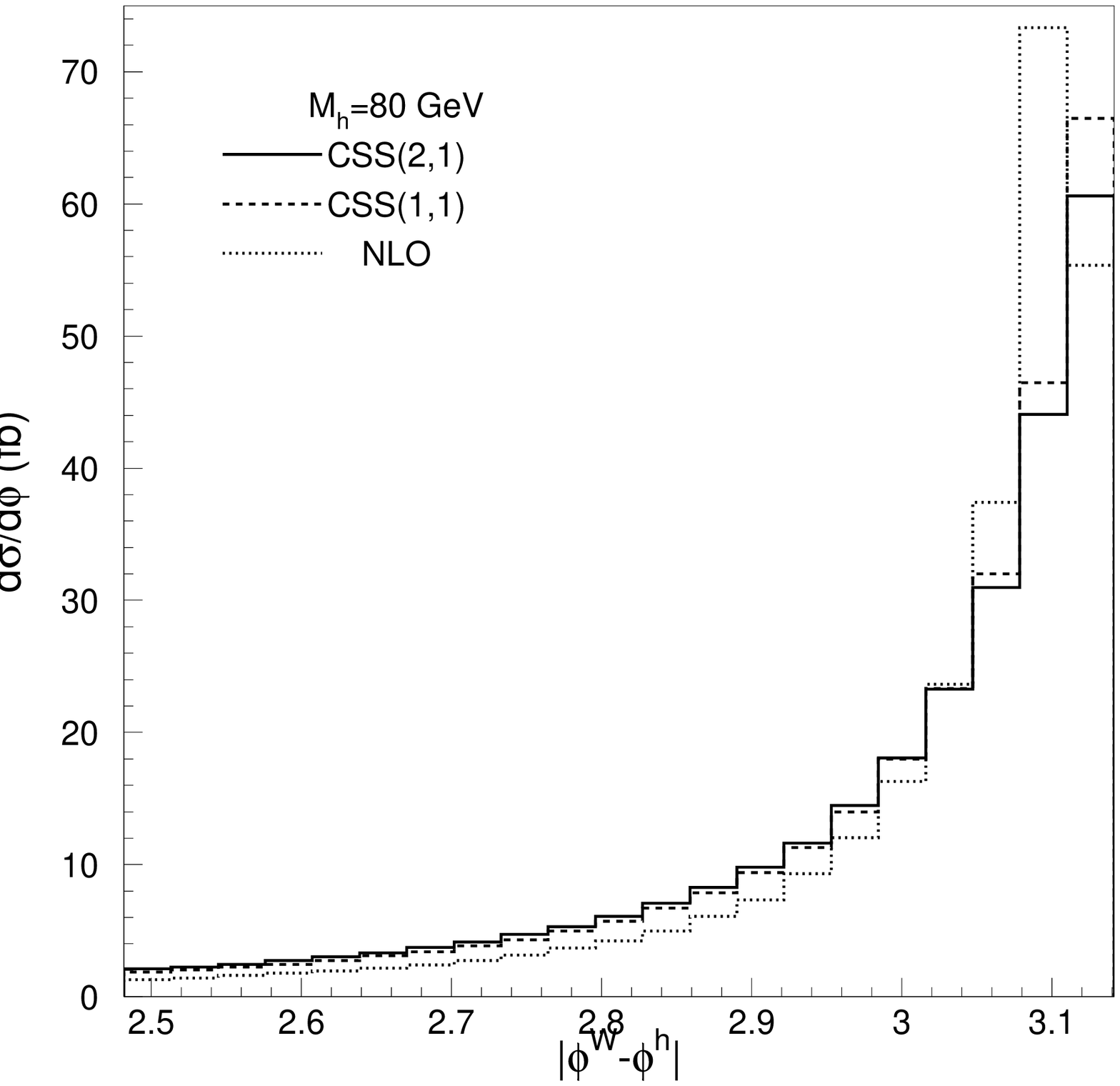}
\vspace*{0in}
\caption{The $\Delta\phi^{W^+h}$ spectrum for $W^+h$ Events}
\label{fig_five}
\end{figure}
\begin{figure}
\centering
\hspace*{0in}
\epsfxsize=3.0in
\epsffile{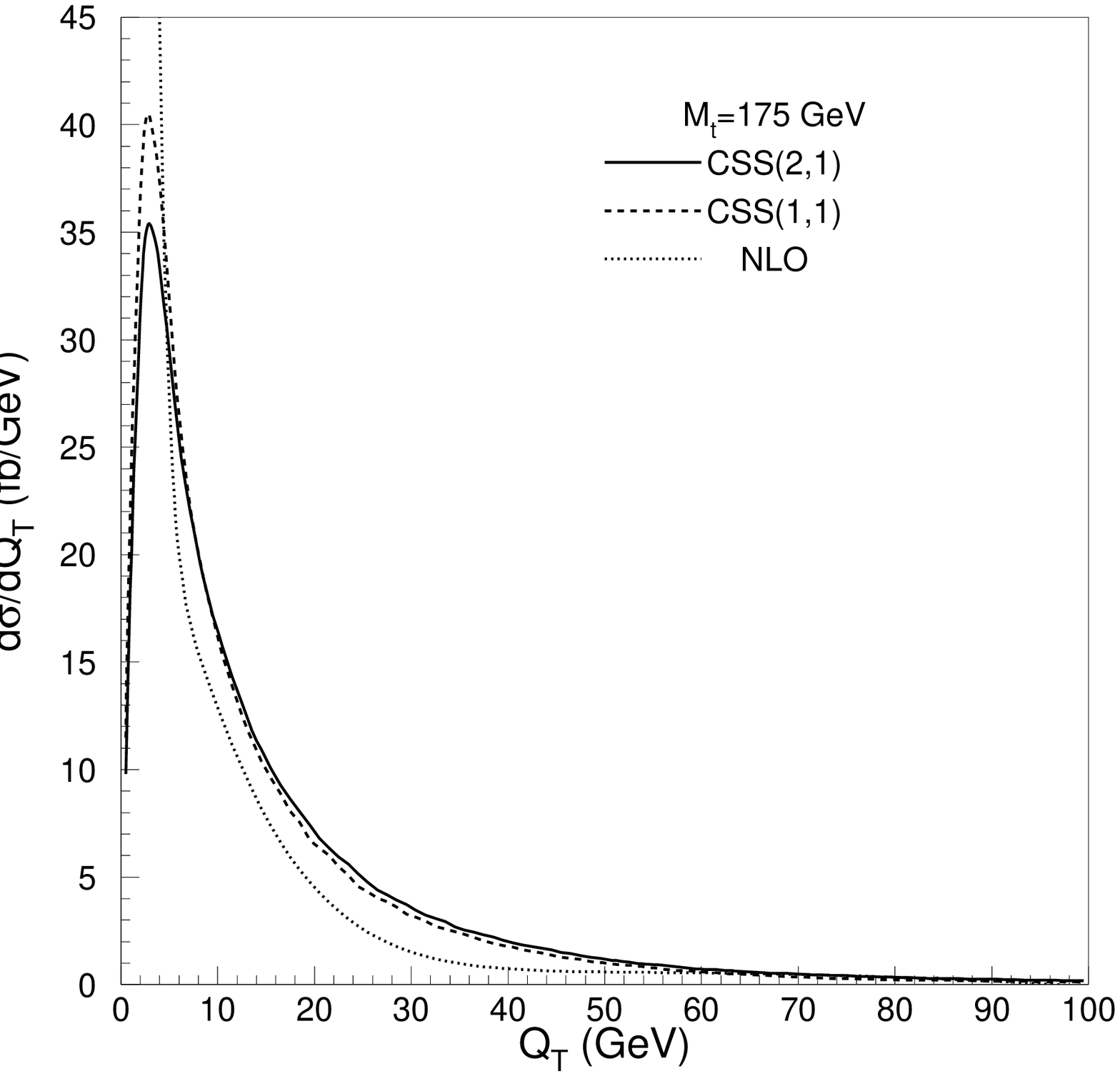}
\vspace*{0in}
\caption{The $Q_T$ spectrum for $t\bar b(g)$ Events}
\label{fig_six}
\end{figure}
\begin{figure}
\centering
\hspace*{0in}
\epsfxsize=3.0in
\epsffile{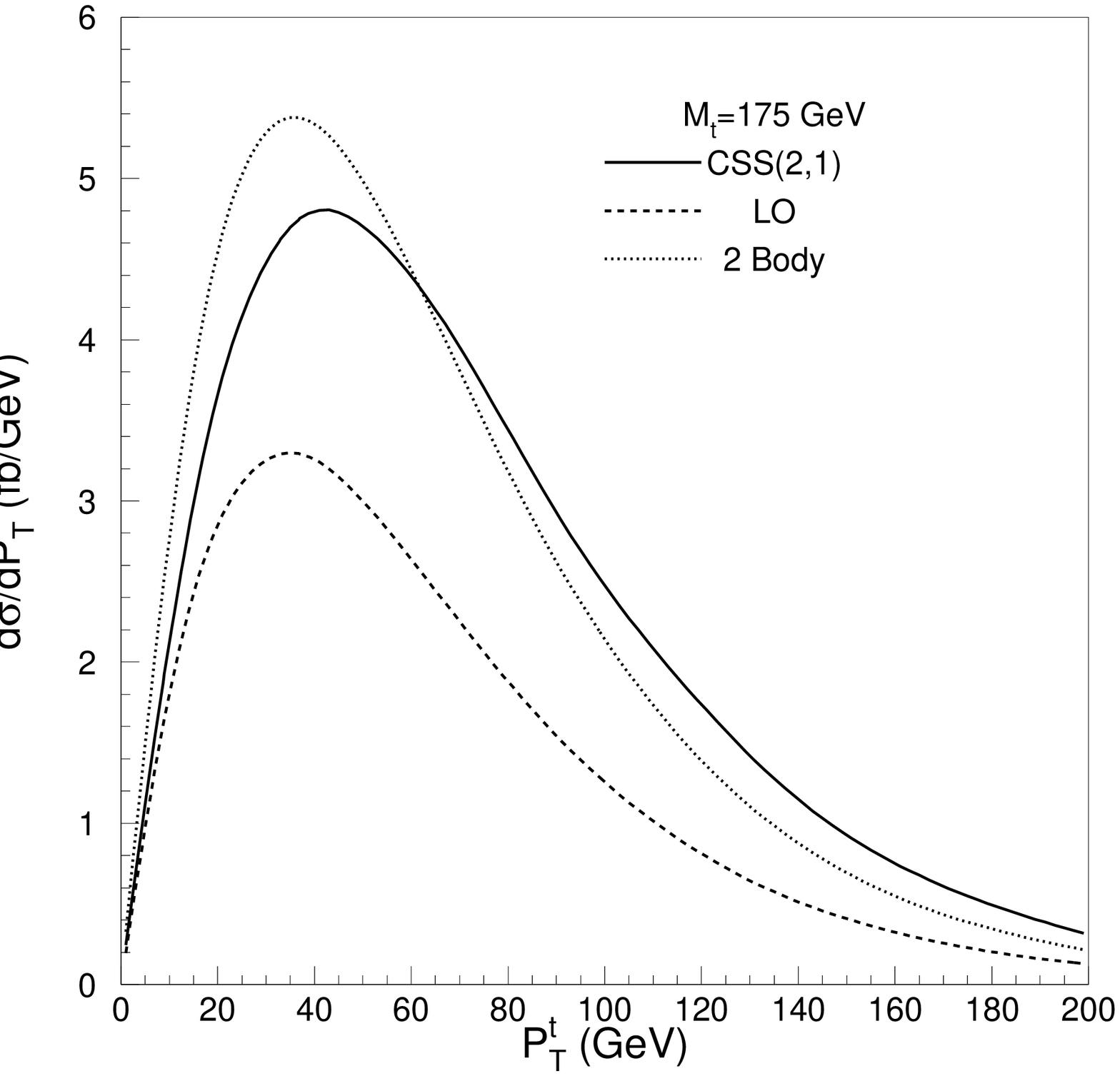}
\vspace*{0in}
\caption{The $P_T^t$ spectra for $t\bar b(g)$ Events}
\label{fig_seven}
\end{figure}
\begin{figure}
\centering
\hspace*{0in}
\epsfxsize=3.0in
\epsffile{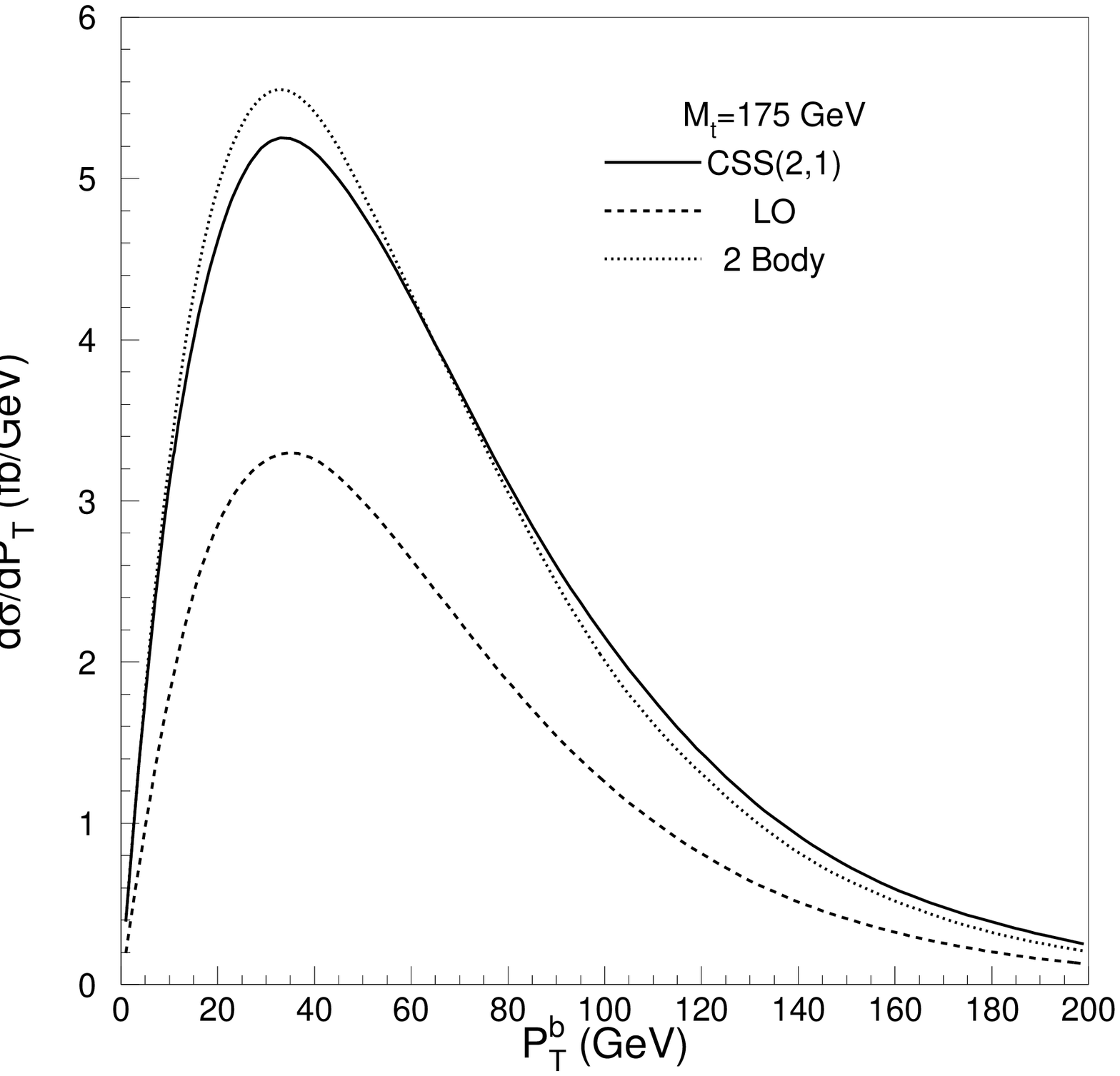}
\vspace*{0in}
\caption{The $P_T^b$ spectra for $t\bar b(g)$ Events}
\label{fig_eight}
\end{figure}
\begin{figure}
\centering
\hspace*{0in}
\epsfxsize=3.0in
\epsffile{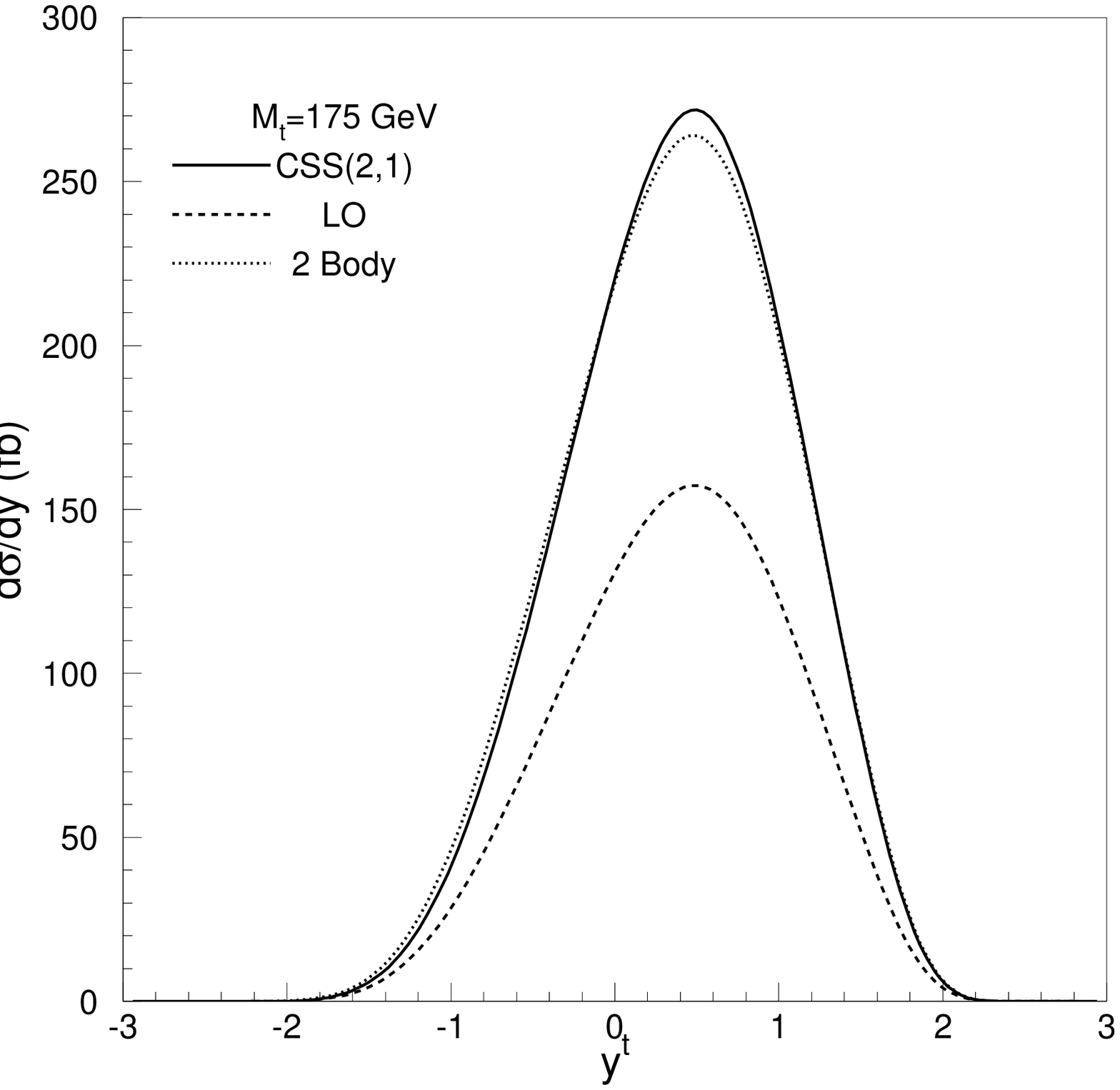}
\vspace*{0in}
\caption{The $y^t$ spectrum for $t\bar b(g)$ Events}
\label{fig_nine}
\end{figure}
\begin{figure}
\centering
\hspace*{0in}
\epsfxsize=3.0in
\epsffile{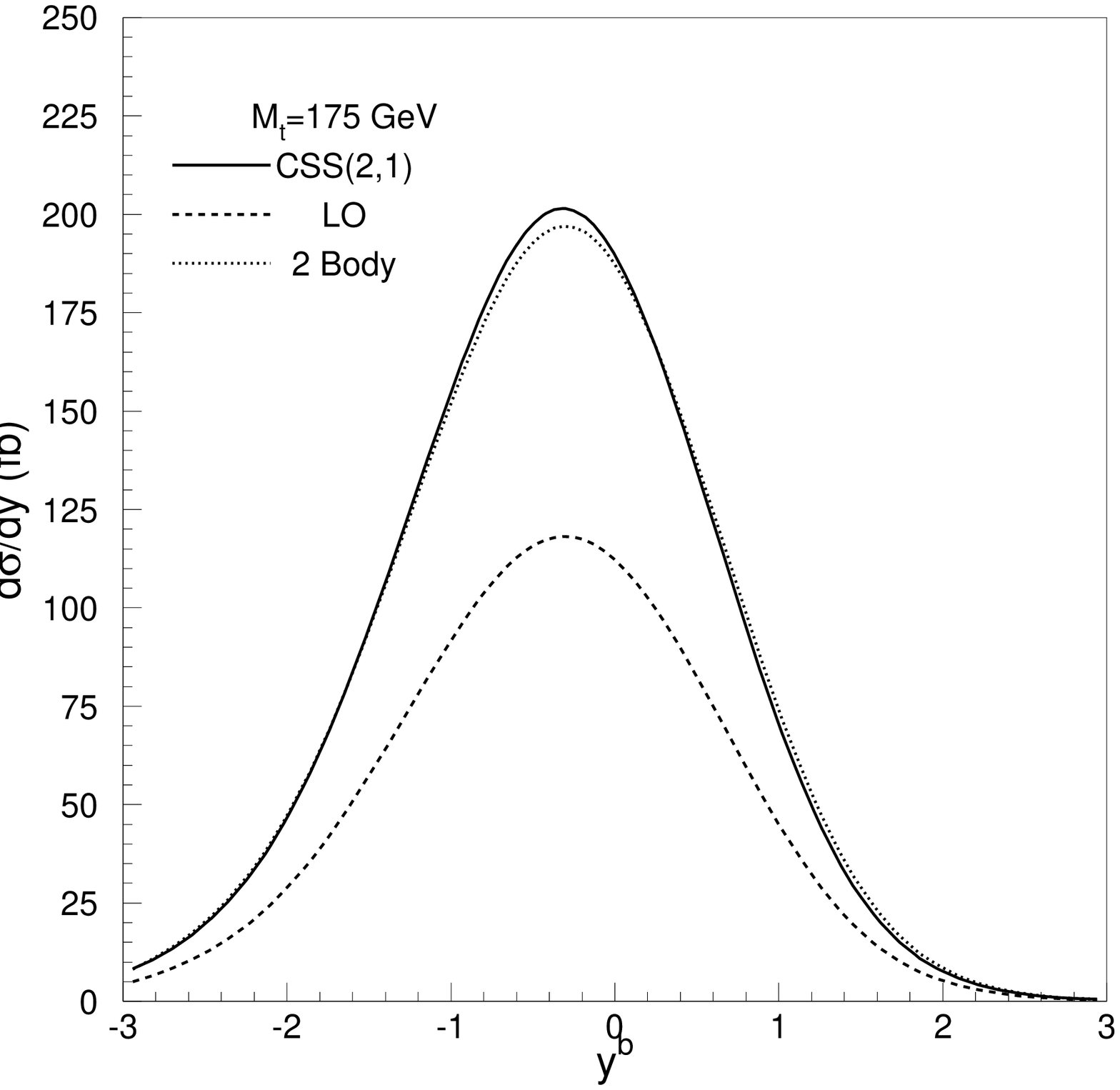}
\vspace*{0in}
\caption{The $y^b$ spectrum for $t\bar b(g)$ Events}
\label{fig_ten}
\end{figure}
\begin{figure}
\centering
\hspace*{0in}
\epsfxsize=3.0in
\epsffile{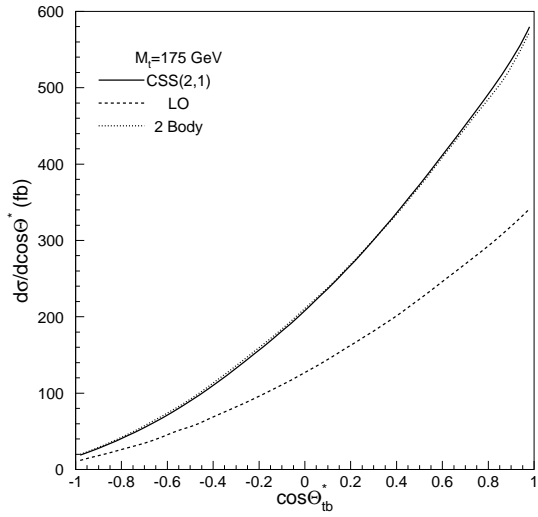}
\vspace*{0in}
\caption{The $\cos\theta^*_{t\bar b}$ spectrum for $t\bar b(g)$ Events}
\label{fig_eleven}
\end{figure}
\begin{figure}
\centering
\hspace*{0in}
\epsfxsize=3.0in
\epsffile{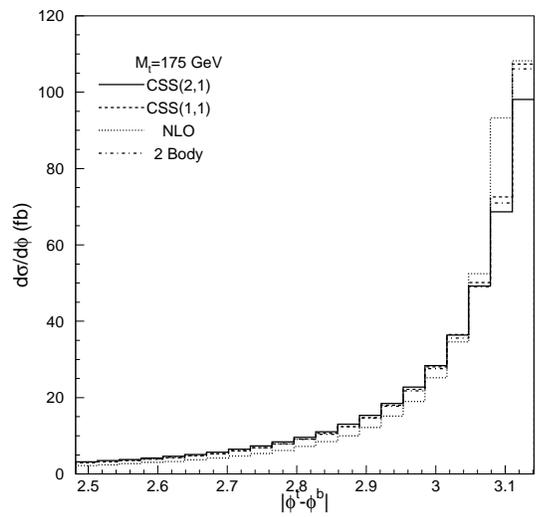}
\vspace*{0in}
\caption{The $\Delta\phi^{t\bar b}$ spectrum for $t\bar b(g)$ Events}
\label{fig_twelve}
\end{figure}
\end{document}